# LARGE SCALE LONGITUDINAL EXPERIMENTS: ESTIMATION AND INFERENCE

APOORVA LAL, ALEX FISCHER, AND MATTHEW WARDROP

ABSTRACT. Large-scale randomized experiments are seldom analyzed using panel regression methods because of computational challenges arising from the presence of millions of nuisance parameters. We leverage Mundlak (1978)'s insight that unit intercepts can be eliminated by using carefully chosen averages of the regressors to rewrite several common estimators in a form that is amenable to weighted-least squares estimation with frequency weights. This renders regressions involving arbitrary strata intercepts tractable with very large datasets, optionally with the key compression step computed out-of-memory in SQL. We demonstrate that these methods yield more precise estimates than other commonly used estimators, and also find that the compression strategy greatly increases computational efficiency. We provide in-memory (`pyfixest`) and out-of-memory (`duckreg`) python libraries to implement these estimators.

# 1. Introduction

Despite the widespread proliferation of experimentation in the technology sector, A/B tests are typically analyzed with simple methods such as t-tests and linear regression adjustment that flatten the time-dimension into a single 'post-treatment' outcome. In the presence of effect heterogeneity across units and over time, the post-treatment average may not be a good summary statistic to feed into decisions. For example, consider the four DGPs illustrated in figure 1a, which all yield the same treatment effect estimate of zero when analyzed using a t-test or a regression on the post-treatment data. These estimates mask considerable heterogeneity: DGP1 contains a true null where the effect is zero for everyone, DGP2 contains a null average effect that masks considerable across-unit heterogeneity, DGP3 contains substantial effect heterogeneity wherein the effect starts off negative but then reverses sign over time (which might correspond to learning effects), and DGP4 contains substantial heterogeneity both over time and across adoption cohorts (where adoption timing may be partially endogenous depending on the frequency with which users interact with a service). Under perfect information, a decision-maker may choose to maintain the status quo under DGP1, personalize treatment under DGP2, and

NETFLIX, TRIVAGO, NETFLIX

run the experiment for longer under DGP3 and DGP4 (with the hope that temporal heterogeneity settles into a steady-state).

Experimenters may want to take temporal heterogeneity seriously for multiple reasons: non-stationarity of treatment effects during the post-treatment window typically suggests that treatments haven't equilibrated into a steady state, and therefore may have limited external validity. Additionally, non-stationarity of treatment effects may also mask evolving demographic heterogeneity; in particular, early estimates may be particularly prone to activity bias (a form of selection-bias wherein initial treatment effects are estimated on highly active users), which may also limit external validity either for rollout or personalized policy learning. Analyzing longitudinal experiments with panel regression techniques readily alleviates these problems, but these methods are used infrequently due to computational challenges arising from a very large number of nuisance parameters.

This paper proposes scalable panel-regression methodologies, along with performant open-source python libraries for both in-memory (`pyfixest`) and out-of-memory (`duckreg`, enabled by tailored compression of the full panel dataset into a set of sufficient statistics) computation. Particular attention is paid to estimating cross-sectional and temporal effect heterogeneity (Abraham et al., 2020; Wooldridge, 2021), uniform inference on the treatment effect function, and omnibus tests for heterogeneous effects to enable specification choice are proposed. We conclude with detailed numerical experiments and runtime comparisons with commonly used specifications and existing software.

## 2. Methods

Consider an experimental setting with N units and T time periods, where a binary treatment $W$ is randomly assigned at time $T_0 < T$, and treatment status is an absorbing state ($W_{it} \geqslant W_{i,t-1}$). This gives rise to two potential outcomes $Y_{it}^0, Y_{it}^1$, and the difference between them in each period $\tau_{it}$ is the individual level dynamic treatment effect. With

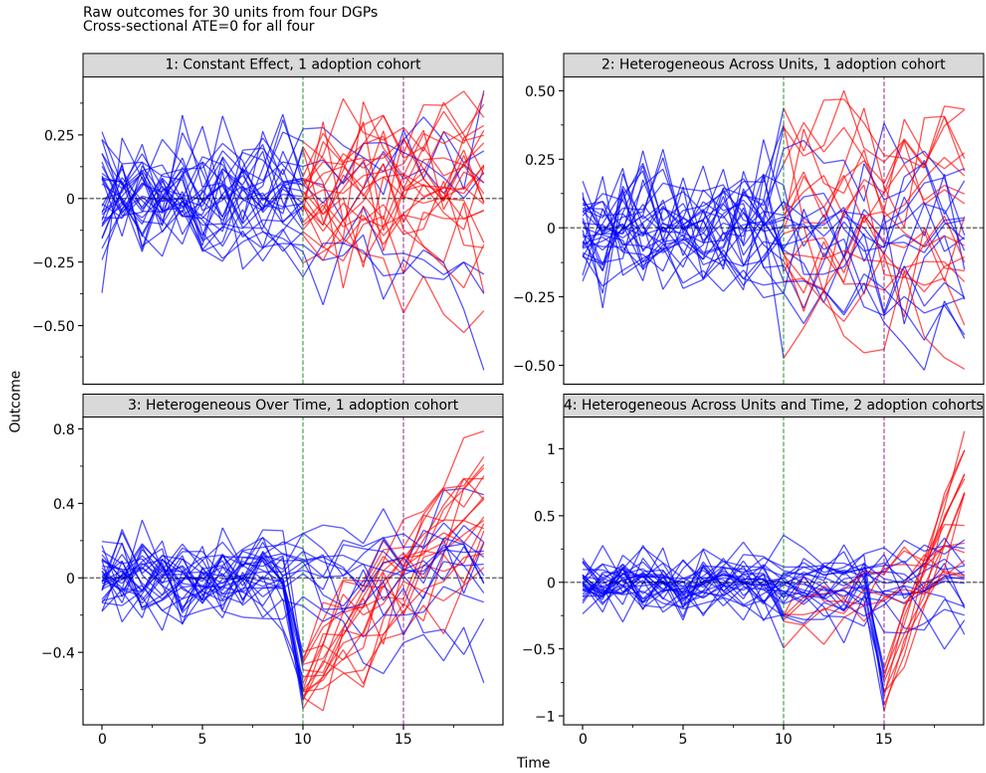

(A) An event-study Anscombe's quartet: all four DGPs yield identical point estimates of 0 when post-randomization data is analyzed using a t-test

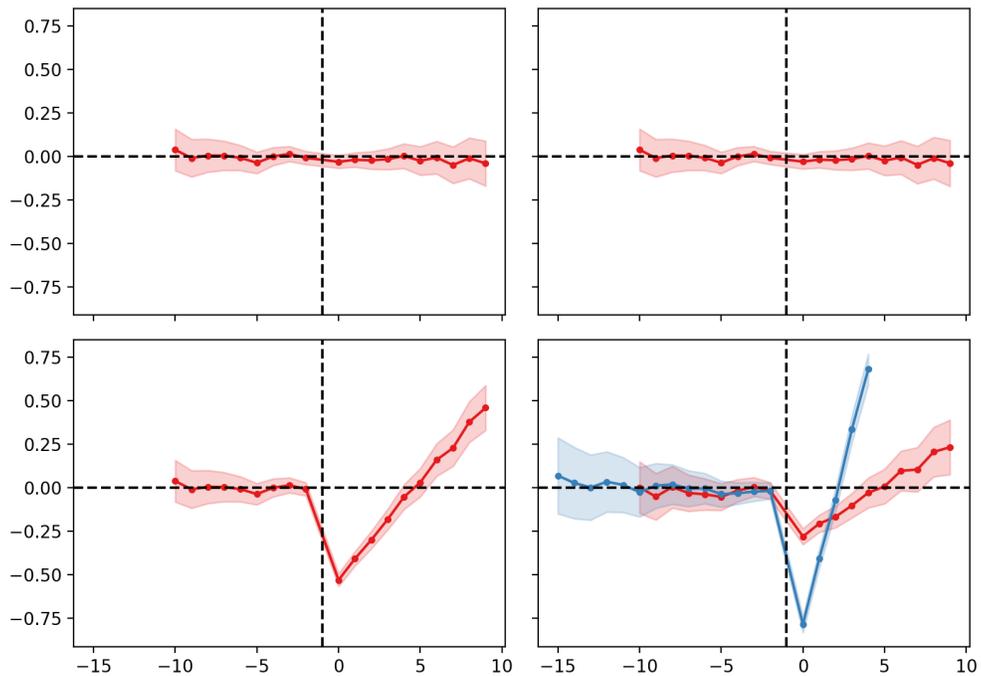

(B) event study (eq 2.6) estimates of DGPs in fig 1a

FIGURE 1. An dynamic-treatment-effect Anscombe's quartet its estimation with event studies

access to this data, an analyst may choose to run one of the following regressions:

$$\overline{Y}_{i,t>T_0} = \alpha + \tau W_i + \varepsilon_i \tag{2.1}$$

$$\overline{Y}_{i,t>T_0} = \alpha + \tau W_i + \beta \overline{Y}_{i,t<T_0} + \varepsilon_i \tag{2.2}$$

$$Y_{it} = \alpha_i + \gamma_t + \tau W_{it} + \varepsilon_{it} \tag{2.3}$$

$$Y_{it} = \alpha + \gamma_t + \sum_{k \geq 0}^{T} \tau_k Z_{it}^k + \varepsilon_{it} \quad \text{where } Z_{it}^k := \mathbb{1}_{\text{argmin}\{t:W_{it}=1\}-t=k} \tag{2.4}$$

$$Y_{it} = \alpha_i + \gamma_t + \sum_{k \neq -1}^{T} \tau_k Z_{it}^k + \varepsilon_{it} \tag{2.5}$$

$$Y_{it} = \alpha_i + \gamma_t + \sum_{c=2}^{C} \sum_{k \neq -1}^{T} \mathbb{1}_{G_i=c} \tau_{kc} Z_{it}^k + \varepsilon_{it} \tag{2.6}$$

All six regressions are well-studied in the econometrics and statistics literature. 2.1 is the simplest difference in means estimator that computes the difference in means for the post-treatment period; 2.2 is a regression adjustment estimator that adjusts for the pre-treatment average outcome $\overline{Y}_{i,t<T_0}$; 2.3 is a panel regression that adjusts for unit and time intercepts but averages treatment effects over post-treatment periods; 2.4 is a dynamic difference in means estimator that computes the difference in average outcomes in each post-treatment period between the treatment and control group by creating separate time-indicators relative to adoption time $Z_{it}$; 2.5 is a panel regression that estimates a separate treatment effect for each time period and conditions on unit intercepts; and 2.6 is an panel regression that estimates a separate treatment effect for each time period for each cohort c defined first treated period (Abraham et al., 2020). 2.6 is equivalent to 2.5 when all treated units adopt the treatment at the same time, but may be meaningfully different under staggered adoption and effect heterogeneity across cohorts, in which case $\tau$ in 2.3 may have the wrong sign relative to all cohort-level effects, which has prompted an explosion of work in applied econometrics (Chaisemartin et al., 2020). All the above specifications are consistent for the cross-sectional ATE under random assignment[1]. However, the first three average over temporal heterogeneity, which may be very important for decisionmaking. 2.3-2.6 are seldom used in large-scale experiemntation frameworks due to computational tractability challenges: typically, they require the use of N × T datasets, which makes in-memory estimation of panel models with industry-scale samples sizes infeasible. On the other hand, efficient estimation of models with unit and time fixed effects requires specialized iterative algorithms, which are difficult to implement in pure SQL (see e.g. Gaure, 2013 for details).

---

[1] Fixed effects regressions estimate $\tau$ consistently even when N→∞ with T fixed, which is a realistic setting in tech experimentation where N ≫ T

## 2.1. Efficient Panel Estimation at Scale.

2.1.1. **Lossless Regression Compression via Sufficient Statistics.** For notational clarity, let us illustrate the core idea of compression on the cross-sectional CUPED regression 2.2, where we regress the outcome on an intercept, the treatment indicator $W_i$, and pre-treatment average $\overline{Y}_{i,t<T_0}$. The simplest approach to estimating this would involve a N × 3 matrix. Alternatively, we can collapse the data to sufficient statistics to strata level, where each stratum j is defined as a unique combination of $(1, W_i, \overline{Y}_{i,t<T_0}) =: \widetilde{\mathbf{X}}_i$. The summary statistics for lossless-compression[2] in a cross-sectional regression involve the outcome sum and sum of squares $\sum_j Y_i, \sum_j Y_i^2 =: \widetilde{y}', \widetilde{y}''$ and the strata size $\sum_j 1(i \in j) =: \widetilde{n}$, as documented by Wong et al. (2021), which can be computed out-of-memory in SQL. With covariates and sufficient statistics stacked in vector form, the coefficient estimates is WLS with frequency weights $\widehat{\beta} = (\widetilde{\mathbf{X}}^\top \text{diag}(\widetilde{\mathbf{n}})\widetilde{\mathbf{X}})^{-1}\widetilde{\mathbf{X}}^\top \text{diag}(\widetilde{\mathbf{n}})\widetilde{\mathbf{y}}$, and the heteroskedasticity-robust Huber-White standard error can constructed with the following sandwich form

$$\widehat{\mathbf{V}}(\beta) = \overbrace{(\widetilde{\mathbf{X}}^\top \text{diag}(\widetilde{\mathbf{n}})\widetilde{\mathbf{X}})^{-1}}^{\text{Bread}} \overbrace{\widetilde{\mathbf{X}}^\top \text{diag}(\sum_j \underbrace{\left(\widehat{y}_j^2 \widetilde{n}_j - 2\widetilde{y}_j \widetilde{y}_j' + \widetilde{y}_j''\right)}_{\text{RSS in j}})\widetilde{\mathbf{X}}}^{\text{Meat}} \overbrace{(\widetilde{\mathbf{X}}^\top \text{diag}(\widetilde{\mathbf{n}})\widetilde{\mathbf{X}})^{-1}}^{\text{Bread}}$$

This compression strategy can drastically reduce the size of the dataset without affecting model estimates, especially when $Y_i$ takes on a discrete set of values: for binary $Y_i$ and a binary treatment $W_i$, the design matrix $1, W_i, \overline{Y}_{i,t<T_0}$ takes on $1 \times 2 \times 2 = 4$ values, so the least squares problem is shrunk from N observations to 4 observations with frequency weights and produces identical point estimates and standard errors.

2.1.2. **The Two-Way Mundlak (TWM) Representation.** For regressions involving two-way fixed effects (2.3-2.6), this approach is ineffective since collapsing the data by $W_{it}, \alpha_i, \gamma_t$ yields no compression at all (since these three jointly constitute a unique observation in the full N × T dataset). This is one possible reason why regressions involving fixed effects are seldom used in very large-scale settings such as online experimentation platforms. However, these regressions can be rendered tractable using an insight that goes back to Mundlak (1978), which is that unit intercepts can be eliminated from a regression by using covariate averages, which also accomodates the use of covariates that are eliminated by the fixed-effects specification (Baltagi, 2023). This approach is expanded to a variety of panel data problems in Wooldridge (2021). Of particular interest is the numerical equivalence (Wooldridge, 2021, thm 3.1) between the two-way fixed-effects coefficient from 2.3-2.6 and the coefficients from the following Two-way Mundlak (TWM) regressions

$$y_{it} = \alpha + \tau W_{it} + \psi \overline{W}_{i,\cdot} + \phi \overline{W}_{\cdot,t} + \varepsilon_{it} \tag{2.7}$$

$$y_{it} = \alpha + \psi D_i + \sum_{k=1}^{T} \phi_t \mathbb{1}_{t=k} + \sum_{k=1}^{T} \tau_k D_i \mathbb{1}_{t=k} + \varepsilon_{it} \tag{2.8}$$

$$y_{it} = \alpha + \underbrace{\sum_{c=1}^{C} \psi_c \mathbb{1}_{D_i=c}}_{\text{Cohort Dummies}} + \underbrace{\sum_{k=1}^{T} \phi_t \mathbb{1}_{t=k}}_{\text{Time Dummies}} + \underbrace{\sum_{c=1}^{C} \sum_{k=1}^{T} \tau_{kc} \mathbb{1}_{D_i=c} \mathbb{1}_{t=k}}_{\text{Cohort} \times \text{Time interactions}} + \varepsilon_{it} \tag{2.9}$$

where we use calendar-time ($t = 1, \ldots, T$) instead of event time (time periods defined relative to first treated period) WLOG for notational simplicity. Unlike eq 2.5, eq 2.7 can be compressed using the strategy outlined in 2.1.1 since the unit and time intercepts have been replaced by unit and time averages of the treatment indicator. To illustrate the compression potential of the Mundlak approach in 2.7, assume that for a balanced experimental panel data set, 50% of units are treated after half the time. For a binary treatment $W_{i,t} \in \{0,1\}$, it follows that $\overline{W}_{i,\cdot} \in \{0, 0.5\}$ (all treated individuals are in the treatment half the time, non-treated individuals are tautologically never treated) and $\overline{W}_{\cdot,t} \in \{0, 0.5\}$ (at any point in time, either 0 or 50% of the population are treated), which leads to $8 = 2 \times 2 \times 2$ potential strata. But because there are no observations with $W_{it} = 1$ but $\overline{W}_{i,\cdot}$ or $\overline{W}_{\cdot,t} = 0$; or observations with $W_{it} = 0$ but $\overline{W}_{i,\cdot} > 0$ and $\overline{W}_{i,\cdot} > 0$, we are left with 4 strata.

---

[2]wherein the compressed representation can recover the coefficients and standard errors estimated on the full uncompressed data (Wong et al., 2021)

Analogous TWM representations for event studies 2.8, 2.9 introduce treatment-cohort dummies $D_i$ and re-introduce time-dummies[3] but eliminate the unit intercepts, which maintains a substantial degree of compression (since $N \gg T$ in settings under consideration) while letting us estimate event studies flexibly under temporal and cross-cohort temporal heterogeneity respectively. The number of observations in a compressed design matrix in 2.8 scales with the number of time periods $2T$, and the number of cohorts $C \cdot T$ in 2.9; both order-of-magnitudes smaller than $N \cdot T$ in 2.3-2.6. We provide a walk-through of the compression approach using a simulated dataset in appendix A.1.

---

[3] Note that we move from event time (where time is normalized relative to the first treatment period) to calendar time $(1, \ldots, T)$ for notational clarity in the Mundlak representation. This is without loss of generality and simplifies computation in SQL.

2.1.2.1. **Standard Errors.** Coefficient estimates from panel regressions are consistent and asymptotically normal based on standard techniques (Wooldridge, 2010). Unlike in the cross-sectional case (2.1.1), we allow residuals to be arbitrarily auto-correlated within units, which motivates the use of cluster-robust standard errors (Bertrand et al., 2004), which are not computable from the compressed data alone. Instead, we may use the cluster-bootstrap or its wild variant wherein we sample units with replacement (MacKinnon, 2023), compress the sampled data, and estimate the coefficient vector repeatedly, which is an embarrassingly parallel procedure and is therefore very scalable. Alternatively, cluster-robust standard-error estimation can be performed by constructing $\Omega = \varepsilon\varepsilon'$ using empirical residuals centrally and parallelizing the estimation of the meat matrix $\sum_i X_i \Omega X_i$ via distributed computing.

2.2. **Testing for the right level of heterogeneity.** Implementing these event studies with linear regression also gives us a straightforward method of testing for the most parsimonious specification. Since 2.7, 2.8 are nested in the most flexible specification 2.9, we can test for across-cohort heterogeneity by conducting Wald tests on the cohort-time coefficients in 2.9, and over-time heterogeneity by conducting a Wald test on the dynamic effects in 2.8. For example, one could test the joint significance of the event study coefficient vector $\{\tau_k\}_{k=1}^T$ in 2.8 against a constant-effects approximation $\tau$ in 2.7, using the latter as a restricted model and the former as the restricted model, and form a test-statistic under the null

$$F_0 = \frac{(RSS - USS)/(T-1)}{USS/(NT-T-1)} \sim F_{T-1, N-T}$$

where RSS and USS are the sum of square residuals under the restricted and unrestricted models respectively. These are aggregate quantities that are easily computable from the compressed data. This approach implicitly follows the standard approach of the test of multiple linear restrictions wherein one forms a $q \times k$ restriction matrix R and tests $R\theta = r$ against $\chi_q^2$ critical values.

# 3. **Numerical Experiments**

3.1. **Estimation Accuracy.** In this section, we perform simulation studies to evaluate the performance of various panel data estimators for treatment effects under different data generating processes (DGPs) that vary cross-sectional and temporal heterogeneity in treatment effects. We consider several functional forms for the treatment effect and assesses the estimators' performance in terms of RMSE[4]. We generate data using the following model: $Y_{it} = \alpha_i + \gamma_t + \beta_i t + \tau_{it} W_{it} + \varepsilon_{it}$, with $i \in [N]$ units observed over $t \in [T]$ time periods where $\alpha_i \sim \mathcal{N}(0, \sigma_\alpha^2)$ is a unit intercept, $\gamma_t \sim \mathcal{N}(0, \sigma_\gamma^2)$ is a time intercept, $\beta_i \sim \mathcal{N}(0, \sigma_\beta^2)$ is a unit time trend, $\tau_{it}$ is a treatment effect with $W_{it}$ randomly assigned at a time period $T_0 + 1 < T$[5], and $\varepsilon_{it}$ is an idiosyncratic error term that is serially correlated within each unit ($\varepsilon_{it} = \rho \varepsilon_{it-1} + \nu_{it}$). We choose this specification due to its flexibility: it maintains random assignment while accommodating substantial unit and temporal heterogeneity in untreated potential outcomes (thanks to the unit intercept and time trend and time intercept), as well as treatment effects. Furthermore, the time-trend component $\beta_i t$ parametrizes the degree of mis-specification: higher values of $\sigma_\beta$ correspond with stronger (unmodeled) unit-level time trends, which and therefore our simulations let us evaluate the robustness of these estimators under a mild but realistic form of mis-specification. The treatment effect $\tau_{it}$ is generated according to one of seven functional forms:

(1) Constant $= \psi \cdot \tau$,
(2) Linear $= \psi \cdot \tau (t - T_0)/(T - T_0)$
(3) (Log) Concave $= \psi \cdot \tau \cdot 0.5 \cdot \log(2(t - T_0 + 1)/(T - T_0) + 1)$
(4) Positive then negative $= \psi \cdot [(t < (T + T_0)/2) \cdot \tau \cdot 2(t - T_0)/(T - T_0) + (t >= (T + T_0)/2) \cdot \tau \cdot (2 - 2(t - T_0)/(T - T_0))]$
(5) Exponential $= \psi \cdot \tau \cdot (1 - \exp(-5(t - T_0)/(T - T_0)))$
(6) Sinusoidal $= \psi \cdot \tau \cdot \sin(2\pi(t - T_0)/(T - T_0))$
(7) Random Walk $= \psi \cdot \tau \cdot \sum_{s=T_0}^{t+1} \eta_s$ ; $\eta_s \sim \mathcal{N}(0, 1)$

Where $\tau$ is a base treatment effect, and $\psi$ is either a scalar (which rules out cross-sectional heterogeneity) or a random variable (which permits both cross-sectional and time-dependent treatment effect heterogeneity). We consider the following estimators:

A. Difference in Means in the post-period (2.1)
B. Regression Adjustment with lagged average (CUPED) (2.2)
C. Static Two-way Fixed Effects (TWFE) (2.3)
D. Difference in means in each time period (2.4)
E. Event Study (2.5)

---

[4] code to reproduce numerical results in this section can be found here
[5] We avoid staggered adoption settings for expositional clarity. In this case, 2.6 collapses into 2.5

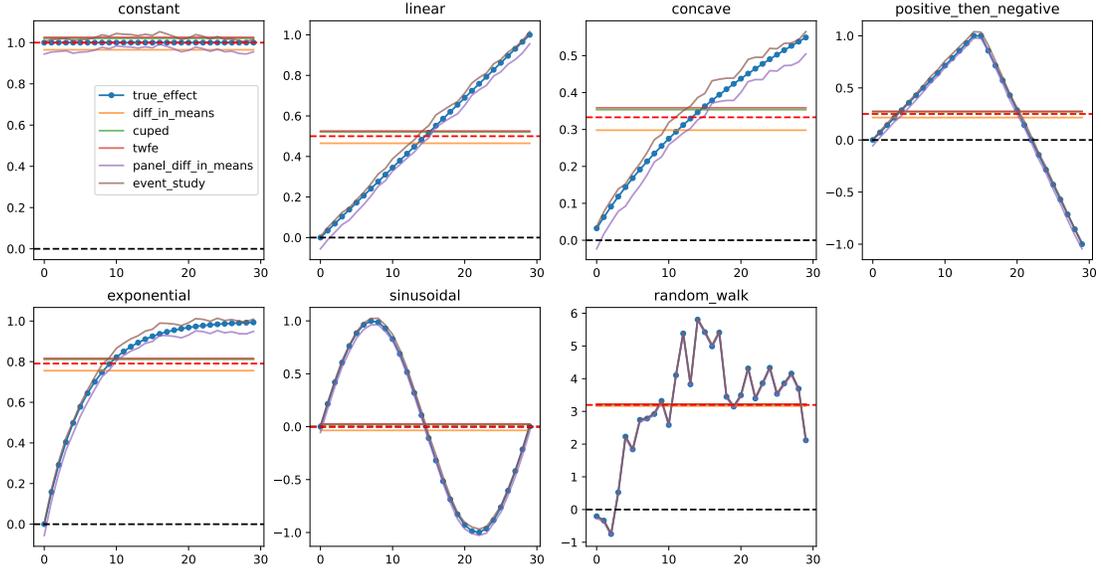

FIGURE 2. Single Realization of considered estimators on different DGPs. True dynamic effect in blue, average of true dynamic effects in red.

We report estimates for all estimators (A-E) for a single draw of each DGP (1-7) in 2. We see that by construction, conventional estimators A-C are static and produce a single number, which is unbiased for the ATE (shown in the red dotted line) but fail to capture treatment effect dynamics. Estimators D-E capture these dynamics accurately, and by incorporating unit-intercepts specification E is more robust than D to confounding or chance imbalances.

We subsequently study the properties of estimators (A-E) on DGPs (1-7) more systematically repeating the procedure in fig 2 1000 times to compute Root Mean Squared Error (RMSE) relative to the true effect functions (1-7) with N = 100,00 units, 35 time periods with treatment randomly assigned with probability 1/2 in periods 15 onwards, and $\sigma_\alpha = 5, \sigma_\gamma = 2, \sigma_\varepsilon = 2, \rho = 0.7$. We vary the presence of unit-level heterogeneity and also vary the variance of unit time trends($\sigma_\beta \in \{0.01, 2\}$, where the values control the degree of mis-specification). We report the RMSE in fig 3 for each estimator under each scenario of unit × time × mis-specification. In all cases with the exception of the constant effects regime, the event study estimator (E) strictly dominates all other estimators. Intuitively, unit level heterogeneity increases RMSE in all cases, as does the degree of mis-specification to a greater degree, without altering the rank ordering of the estimators. While the gap in RMSE between the event study and dynamic difference in means estimators is relatively small, the latter is also quite sensitive to confounding and is only consistent under randomization, while the former is robust to confounding and retains consistency under parallel trends, which is an additional reason to prefer it.

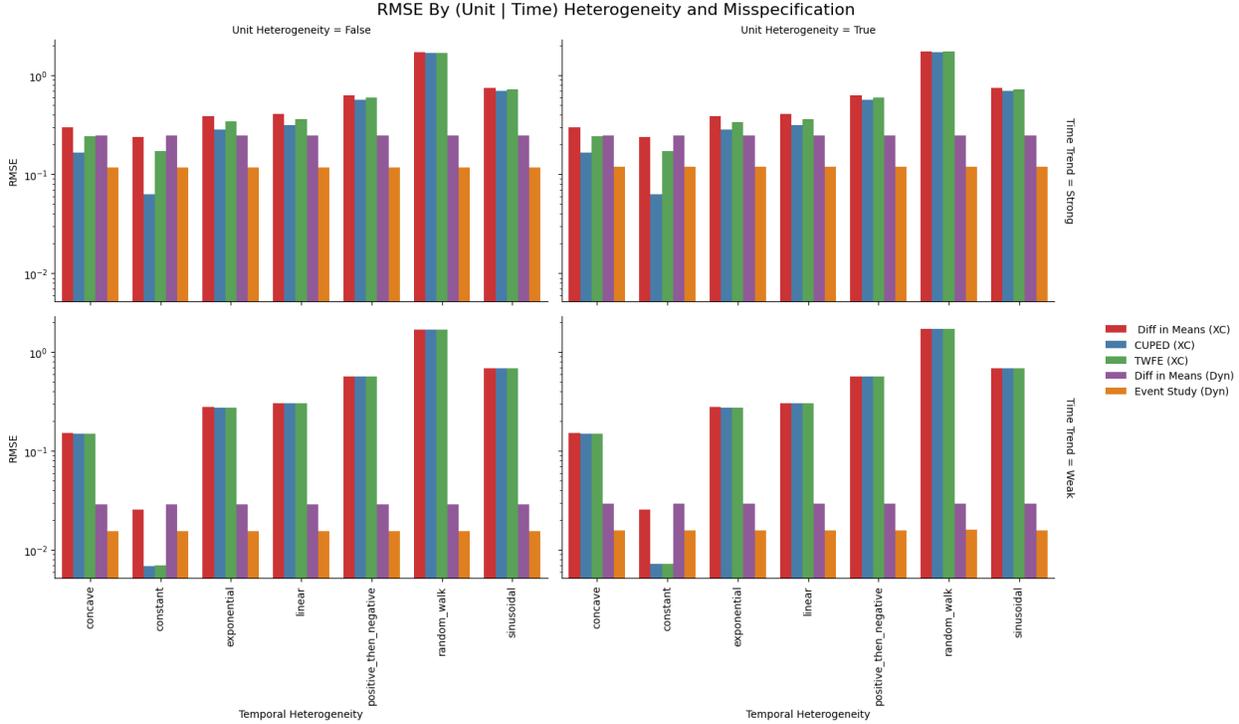

FIGURE 3. RMSE of Estimators (A-E) under temporal heterogeneity (1-7), cross-sectional heterogeneity (T/F) and degree of mis-specification(T/F).

3.2. **Timing.** We provide timing benchmarks of the mundlak-and-compression strategy for both the two-way fixed effects estimator (equation 2.3) and event study estimator (equation 2.5) provided via the duckreg and pyfixest libraries. All benchmarks are conducted on simulated panel data sets with increasing number of units - from one thousand to ten million - and increasing time window - from 14 to 42 time periods (equivalent to two / six weeks of repeated observations). As a result, the smallest test data set contains $1000 \times 14 = 14,000$ observations, while the largest consists of a sample of 140 million observations. For each data set and model, we report the median runtime out of three repetitions. All benchmarks were run using Python 3.12.7 on an Intel 64-bit 8-core processor with 16GB of RAM. Code to replicate the benchmarks is hosted on github.[6]

---

[6]Code for running all benchmarks can be found at https://github.com/py-econometrics/panel-at-scale-code.

Table 1 summarizes the runtime for the two-way fixed effects estimator 2.3. As the benchmarks aim to capture the runtime of fitting model coefficients, but not to conduct inference, we choose the minimum supported number of bootstrap repetitions for all timings.[7] Starting from small sample sizes, duckreg's implementation is orders of magnitudes faster than pyfixest' in memory implementation of the mundlak-and-compression strategy and pyfixest's fit on the uncompressed data. The out-of-memory implementation via duckdb in duckreg particularly shines for large data sets - its median runtime with 10 million treated units observed over 14 periods is around 20 seconds, while pyfixest takes over 2 minutes on the long data. We further benchmark against the statsmodels library and find that for the smallest data sets, statsmodels is around 90x slower than duckreg and 15x slower than pyfixest, which is why we decided to not include further benchmarks on larger data sets for statsmodels.[8]

TABLE 1. TWFE Benchmarks

| Observations | Units | Periods | duckreg | pyfixest | pyfixest compressed | statsmodels |
|---|---|---|---|---|---|---|
| 14K | 1K | 14 | 0.03 | 0.18 | 0.17 | 2.75 |
| 28K | 1K | 28 | 0.03 | 0.19 | 0.18 | 6.27 |
| 42K | 1K | 42 | 0.04 | 0.20 | 0.21 | 9.43 |
| 140K | 10K | 14 | 0.05 | 0.22 | 0.26 | x |
| 280K | 10K | 28 | 0.06 | 0.26 | 0.36 | x |
| 420K | 10K | 42 | 0.04 | 0.31 | 0.47 | x |
| 1M | 100K | 14 | 0.07 | 0.64 | 1.27 | x |
| 3M | 100K | 28 | 0.21 | 1.00 | 2.28 | x |
| 4M | 100K | 42 | 0.24 | 1.41 | 3.43 | x |
| 14M | 1M | 14 | 1.03 | 4.88 | 22.12 | x |
| 28M | 1M | 28 | 3.41 | 8.07 | 62.07 | x |
| 42M | 1M | 42 | 10.92 | 13.60 | 117.63 | x |
| 140M | 10M | 14 | 19.70 | 123.88 | x | x |
| Mundlak | | | ✓ | - | ✓ | - |
| Compression | | | ✓ | - | ✓ | - |
| Out-of-Memory | | | ✓ | - | - | - |

All benchmarks are measured in seconds and report the median runtime out of three iterations.

Table 2 summarizes benchmarks for the event study model in equation 2.5. We do not include benchmarks for in-memory mundlak-and-compress via pyfixest. For statsmodels, we only provide reference benchmarks for the smallest benchmark data set. As before, duckreg's compressed implementation beats pyfixest's in memory implementation on the uncompressed data set. Once again, statsmodels is orders of magnitude slower than both alternatives for the smallest data sets.

---

[7]For all regressions via duckreg, this implies that we set the number of bootstrap iterations to 0. For all regressions through pyfixest using the Mundlak transform, we set the number of bootstrap repetitions to 1. Note that pyfixest and statsmodels always compute inference when fitting a model on uncompressed data.
[8]We stopped the benchmark for statsmodels with 10K units after approximately 30 minutes. For all data sets with 100K units, statsmodels fails with an out-of-memory error. We have not attempted to fit models on more than 100K units.

TABLE 2. Event Study Benchmarks

| Observations | Units (N) | Time Periods (T) | duckreg | pyfixest | statsmodels |
|---:|---:|---:|---:|---:|---:|
| 14K | 1K | 14 | 0.07 | 0.22 | 3.45 |
| 28K | 1K | 28 | 0.14 | 0.26 | 7.61 |
| 42K | 1K | 42 | 0.23 | 0.36 | 12.07 |
| 140K | 10K | 14 | 0.24 | 0.35 | x |
| 280K | 10K | 28 | 0.71 | 0.81 | x |
| 420K | 10K | 42 | 1.02 | 1.39 | x |
| 1M | 100K | 14 | 1.39 | 1.63 | x |
| 3M | 100K | 28 | 6.75 | 5.75 | x |
| 4M | 100K | 42 | 11.78 | 16.25 | x |
| 14M | 1M | 14 | 17.01 | 22.77 | x |
| 28M | 1M | 28 | 84.48 | 289.28 | x |
| 42M | 1M | 42 | 292.15 | x | x |
| 140M | 10M | 14 | 644.86 | x | x |
| Mundlak | | | ✓ | - | - |
| Compression | | | ✓ | - | - |
| Out-of-Memory | | | ✓ | - | - |

All benchmarks are measured in seconds and report the median runtime out of three iterations.

# 4. Conclusion

We emphasize the untapped utility of taking treatment effect dynamics seriously, and propose a solution in the analysis of large scale online experiments. Despite the existence of well-known estimators from the panel data literature to estimate dynamic treatment effects in such settings, take-up in industrial applications has been weak, largely due to computational constraints arising from the presence of unit intercepts. We propose a solution to this problem by pairing the Mundlak representation with compressed regression with sufficient statistics. Finally, we conclude with comprehensive numerical experiments that illustrate that panel data estimators substantially out-perform the popular estimators in the presence of temporal heterogeneity in treatment effects. While we focus on the most common use case of absorbing treatments (wherein units that are assigned to treatment remain treated in all subsequent periods) in the present work, analyzing settings where units adopt sequences of treatments remains challenging from both a theoretical and computational perspective, and is a direction for future work.

# Appendix A. Computational Examples

## A.1. **Example of Compressed Regression using** `duckreg`.

A.1.1. **One-shot Adoption.** We simulate a panel dataset with the following characteristics:

- 500,000 units observed over 30 time periods
- 250,000 units receive treatment starting at period 15
- Treatment effect follows a "sharkfin" pattern: increasing logarithmically for 8 periods post-treatment, then returning to zero
- Unit and time fixed effects are included, with autoregressive error structures

The treatment assignment is visualized in Figure 4.

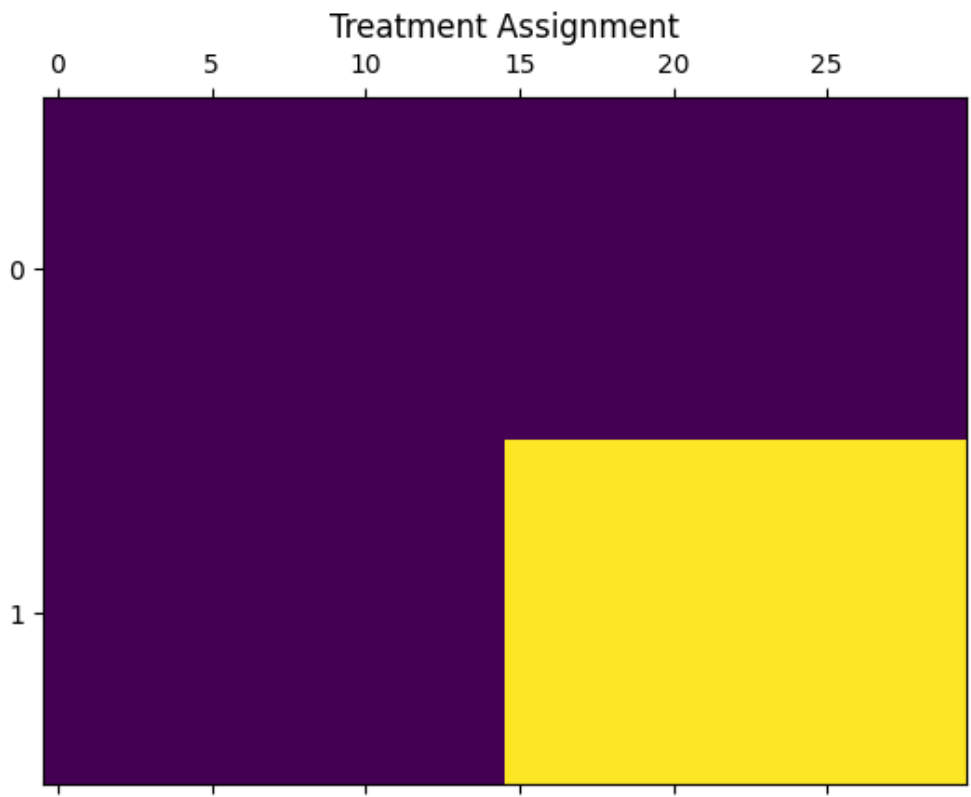

FIGURE 4. Treatment Assignment

We compare three estimation approaches:

(1) Two-way fixed effects (2WFE)

(2) Static two-way Mundlak
(3) Dynamic two-way Mundlak (Event Study)

The cross-sectional results are summarized in table 3, and the dynamic results are summarized in 5

TABLE 3. Estimation Results

| Method | Estimate | Standard Error |
|---|---|---|
| True Average Effect | 0.2153 | - |
| 2WFE | 0.2152 | 0.0006 |
| Static Mundlak | 0.2152 | 0.0004 |

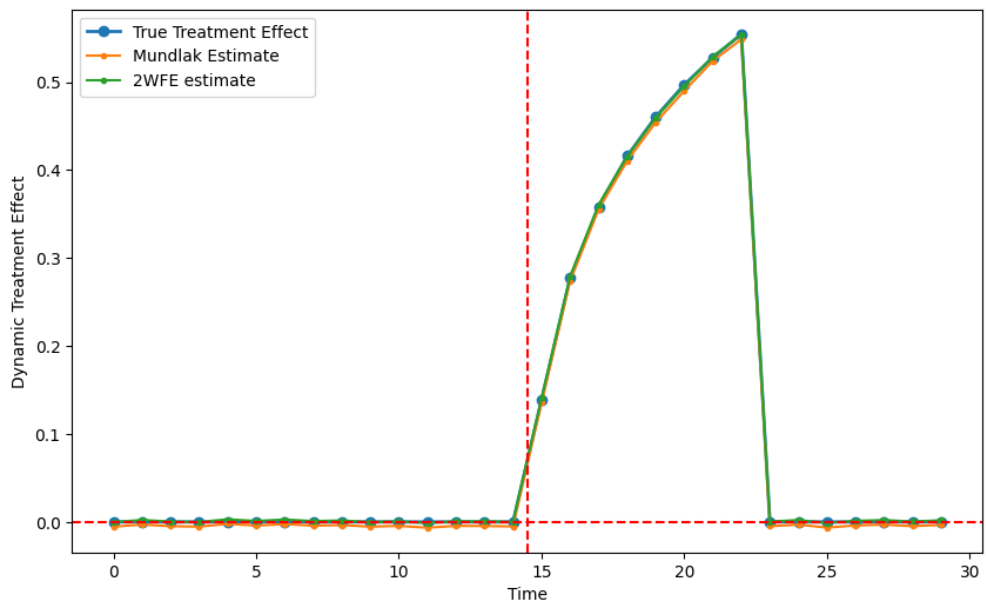

FIGURE 5. Estimated Treatment Effects

As shown in Figure 5, both the two-way fixed effects and Mundlak approaches accurately recover the true treatment effect pattern.

A.1.1.1. **Implementation.** The compressed Mundlak regression can be implemented using the following Python code:

```
mundlak = DuckMundlakEventStudy(
    db_name="database.db",
    table_name="panel_data",
    outcome_var="Y_it",
    treatment_col="W_it",
```

```
    unit_col="unit_id",
    time_col="time_id",
    cluster_col="unit_id",
    seed=42,
    pre_treat_interactions=True,
)

mundlak.fit()
evsum = mundlak.summary()
```

This approach significantly reduces computational overhead, compressing the original dataset of 15,000,000 observations to just $C \times T = 2 \times 30 = 60$ rows for estimation.

### A.1.2. Staggered Adoption.

- Multiple cohorts of varying cohort sizes receive treatment at different time points
- Each cohort has a unique treatment effect pattern
    - Cohort 1: mean reversal: big bump that decays to zero within 10 days, then zero
    - Cohort 2: shark-fin - logarithmic for the first week, then 0
    - Cohort 3: sinusoidal
- Treatment effects vary in magnitude and duration across cohorts

The same `duckreg` function call as above can be used to estimate treatment effects in the staggered setting.

Treatment adoption timing and resultant estimates are summarized in fig 6. This demonstrates the flexibility and accuracy of the Compressed Mundlak Regression approach in handling both simple (one-shot) and complex (staggered) treatment adoption scenarios in large panel datasets.

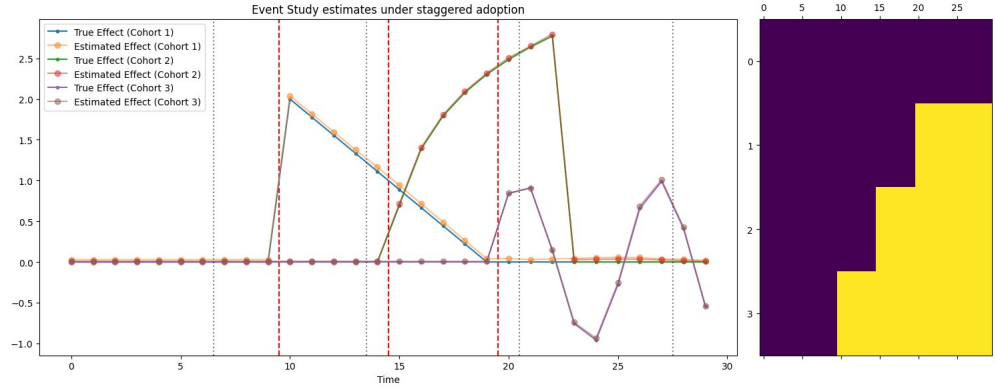

FIGURE 6. Adoption and Estimates under staggered adoption with a never treated group